\documentclass[conference]{IEEEtran}
\IEEEoverridecommandlockouts
\usepackage{setspace}
\usepackage{cite}
\usepackage{amsmath,amssymb,amsfonts}
\usepackage{multirow}
\usepackage[linesnumbered,ruled]{algorithm2e}
\usepackage{array}
\usepackage{url}
\usepackage{verbatim}
\usepackage{stfloats}
\usepackage{graphicx}
\usepackage{subfigure}
\usepackage{float}
\usepackage{textcomp}
\usepackage{xcolor}
\usepackage{soul}
\usepackage{booktabs}
\usepackage{bm}
\def\BibTeX{{\rm B\kern-.05em{\sc i\kern-.025em b}\kern-.08em
		T\kern-.1667em\lower.7ex\hbox{E}\kern-.125emX}}
\usepackage{balance}
\definecolor{matlabyellow}{rgb}{0.9290,0.6940,0.1250}
\begin{document}
	\sethlcolor{yellow} 
	
	\title{PINN and GNN-based RF Map Construction for Wireless Communication Systems}

	\author{
		\IEEEauthorblockN{Lizhou Liu$^1$, Xiaohui Chen$^1$, Zihan Tang$^2$, Mengyao Ma$^2$, Wenyi Zhang$^{1*}$}

		\IEEEauthorblockA{$^1$Department of Electronic Engineering and Information Science,\\
		University of Science and Technology of China, Hefei, Anhui 230027, China\\
		$^2$Wireless Technology Lab, Huawei, Shenzhen 518129, China\\
		Email: liulizhou@mail.ustc.edu.cn; cxh@ustc.edu.cn; \{tangzihan1; ma.mengyao\}@huawei.com; wenyizha@ustc.edu.cn}

		\vspace{-1cm}

		\thanks{This work was supported by the National Natural Science Foundation of China under Grant No. 62231022. Corresponding author: Wenyi Zhang.}
	}
	
	\maketitle

	\begin{abstract}
		Radio frequency (RF) map is a promising technique for capturing the characteristics of multipath signal propagation, offering critical support for channel modeling, coverage analysis, and beamforming in wireless communication networks. This paper proposes a novel RF map construction method based on a combination of physics-informed neural network (PINN) and graph neural network (GNN). The PINN incorporates physical constraints derived from electromagnetic propagation laws to guide the learning process, while the GNN models spatial correlations among receiver locations. By parameterizing multipath signals into received power, delay, and angle of arrival (AoA), and integrating both physical priors and spatial dependencies, the proposed method achieves accurate prediction of multipath parameters. Experimental results demonstrate that the method enables high-precision RF map construction under sparse sampling conditions and delivers robust performance in both indoor and complex outdoor environments, outperforming baseline methods in terms of generalization and accuracy.
	\end{abstract}
	

	\section{Introduction}
	
	The rapid advancement of emerging applications such as the Internet of things, autonomous vehicles, and smart cities has increased demand for highly efficient and reliable wireless communication systems [1].
	Precise characterization of radio frequency (RF) signal distribution and propagation in target areas is essential for extracting and exploiting channel state information (CSI).
	High-precision RF maps effectively analyze and predict the spatial distribution of RF signals, which is crucial for enhancing communication system quality of service (QoS). 

	A key challenge in using RF maps to enhance communication system performance is constructing complete and accurate maps.
	Traditional RF map construction relies on empirical propagation models, such as the Okumura-Hata model \cite{Hata} and the 3GPP urban macro model \cite{3GPP}. 
	However, these models have limited capability in modeling complex geographical environments and often fail to capture real-world propagation scenarios accurately.
	Ray tracing offers an alternative approach by simulating electromagnetic wave propagation \cite{Sionna}. Although it generates more accurate RF maps, simulating millions of rays demands substantial computational resources and results in significant time overhead.
	Interpolation-based methods (e.g., Kriging \cite{Kriging}) offer improved computational efficiency when using sparse data samples. However, their strong dependence on the spatial distribution of data points often results in significant estimation errors in sparse or non-uniform sampling scenarios.

	Advances in artificial intelligence (AI) technologies have opened new opportunities for RF map construction. RadioUNet \cite{UNet} pioneered deep learning-based path loss map generation, while a cascaded U-Net model \cite{WNet} predicted signal strength distribution using environmental data and sparse sampling. Diffusion models have also been applied to generative RF map construction \cite{diffusion}. These approaches primarily predict single physical quantities, such as path loss or signal strength, with only the recent work \cite{DoD} attempting to generate angle of arrival (AoA) and angle of departure (AoD) maps. Meanwhile, most existing methods focus solely on the strongest arriving path and require a large amount of training data, thus limiting their applicability in complex environment.

	Since multipath propagation dominates RF signal transmission, constructing the multipath map is crucial for accurate signal modeling \cite{use1}. 
	However, due to the complexity of propagation and the interdependence of multipath parameters, many general-purpose AI methods cannot directly capture these relationships, particularly under sparse data. Physics-informed neural networks (PINN) \cite{PINN1} incorporate physical laws into machine learning to ensure data fitting while satisfying physical constraints, and have proven effective in complex systems like fluid dynamics. Given the inherent physical constraints in our task, we adopt PINN to ensure physically consistent predictions. To further capture spatial dependencies in RF signal propagation, graph neural networks (GNN) \cite{GNN1} have been employed to model topological relationships and hence can be beneficial for multipath inference.

	This paper proposes a novel RF map construction method that combines PINN and GNN. Our approach first decomposes multipath signals into key physical components, including received power, time delay, and AoA (both elevation and azimuth angles) for each path. The PINN computes theoretical predictions by incorporating electromagnetic propagation constraints and intrinsic relationships among these components, while the GNN captures spatial correlations to refine these predictions. 
	We conducted experiments in two scenarios: an indoor environment using the public DeepMIMO dataset \cite{DeepMIMO} and a campus environment modeling the West Campus of the University of Science and Technology of China (USTC). The experimental results demonstrate that our method performs exceptionally well in sparse RF map construction for both scenarios. Compared to separately estimating individual physical components or using Kriging, standalone GNN, or PINN approaches, our method significantly improves prediction accuracy across all physical components.

	\section{Multipath System Model}
	
	\addtolength{\topmargin}{0.019in}

	In wireless signal propagation, information is encoded in both amplitude and phase of electromagnetic waves. In typical environments, transmitted signals reach the receiver through multiple paths, including line-of-sight (LoS) and non-LoS (NLoS) paths caused by reflection, scattering, and diffraction. This propagation process strictly follows Maxwell's equations. 
	When waves interact with environmental scatterers having different dielectric properties, they produce multipath components with varying delay, direction, and fading characteristics. The superposition of these multipath components in both temporal and spatial domains creates the received signal, a phenomenon formally known as multipath propagation.

	Consider a wideband wireless communication system where a transmitter sends a baseband equivalent signal $s(t)$. For a receiver equipped with a uniform planar array (UPA) consisting of $M \times N$ antennas, the received signal $\mathbf{r}(t)$ under multipath propagation is given by
	\vspace{-0.1cm}
	\begin{equation}
		\mbox{\small$\displaystyle
		\mathbf{r}(t) = \sum_{l=1}^{L} \alpha_l \mathbf{a}(\theta_l, \varphi_l) s(t - \tau_l) + \mathbf{n}(t),
		$}
		\label{r(t)}
	\end{equation}
	where $L$ is the number of multipath components, $\alpha_l = |\alpha_l| e^{-j2\pi f_c \tau_l}$ is the complex gain of the $l$-th path, with $f_c$ denoting the carrier frequency and $\tau_l$ the corresponding path delay, $\theta_l \in [0,\pi]$ and $\varphi_l \in [-\pi, \pi]$ represent the elevation and azimuth angles of the $l$-th path, respectively, and $\mathbf{n}(t)$ denotes the additive white Gaussian noise (AWGN) vector. 
	For a UPA placed on the $x-y$ plane with antenna spacings $d_x$ and $d_y$, the array response vector $\mathbf{a}(\theta_l, \varphi_l) \in \mathbb{C}^{MN \times 1}$ is expressed as
	\begin{equation}
		\mbox{\small$\displaystyle
		\mathbf{a}(\theta_l, \varphi_l) = \left[e^{-j \frac{2\pi}{\lambda} (m d_x \sin \theta_l \cos \varphi_l + n d_y \sin \theta_l \sin \varphi_l)}\right]_{MN \times 1},
		$}
		\label{a}
	\end{equation}
	where $\lambda= {c}/{f_c}$ is the wavelength of the transmitted signal, $c$ is the speed of light, and $m \in \left\{0, 1, \dots, M-1\right\}$ and $n \in \left\{0, 1, \dots, N-1\right\}$ are antenna indices along the $x$- and $y$-axes of the UPA, respectively.

	The time-domain channel impulse response $\mathbf{h}(t)$ is then written as
	\begin{equation}
		\vspace{-0.1cm}
		\mbox{\small$\displaystyle
		\mathbf{h}(t)=\sum_{l=1}^{L} |\alpha_l| e^{-j2\pi f_c \tau_l} \mathbf{a}(\theta_l, \varphi_l) \delta(t - \tau_l),
		$}
		\label{h(t)}
		\vspace{-0.1cm}
	\end{equation}
	where $\delta(t)$ denotes the Dirac delta function.

	According to \eqref{r(t)}-\eqref{h(t)}, the multipath-related physical quantities in wireless communication systems are parameterized into $\alpha_l$, $\tau_l$, $\theta_l$, and $\varphi_l$. Since $\alpha_l$ is a complex quantity that is expressed using $\tau_l$ and the received signal power $p_l =|\alpha_l|^2$ of the $l$-th path, i.e., $\alpha_l = \sqrt{p_l}e^{-j2\pi f_c \tau_l}$, the multipath channel is fully described by the quadruple parameters $\left\{ p_l, \tau_l, \theta_l, \varphi_l\right\}_{l=1}^L$. These parameters describe the spatial coverage and angular properties of the signal, enabling direct support for antenna configuration, beamforming, and localization applications.

	\section{RF Map Construction via PINN and GNN}
	In this section, we propose a PINN-GNN joint framework for RF map construction, which predicts the multipath parameters across the entire region based on sparsely sampled measurements at known locations. 
	\begin{figure*}[htbp]
		\centering
		\includegraphics[width=0.937\textwidth]{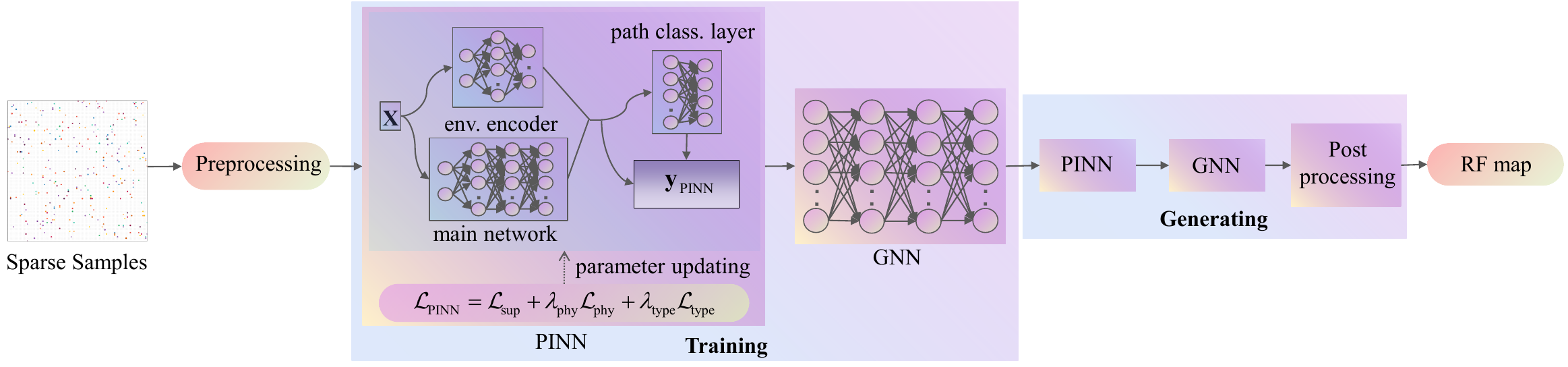}
		\vspace{-0.26cm}
		\caption{PINN-GNN architecture for RF map construction.}
		\label{fig:PINN_GNN}
		\vspace{-0.5cm}
	\end{figure*}

	\subsection{PINN Architecture}

	The PINN is designed to learn the mapping from spatial locations to multipath parameters under the guidance of electromagnetic propagation rules. It consists of three main components: an environment encoder, a main network, and a path classification layer. The model takes spatial coordinates of known samples as inputs and uses their multipath data for supervision, generating predictions at all target locations.

	\textit{\textbf{Environment Encoder}}: The environment encoder uses a three-layer fully connected network. It takes the 2-dimensional (2D) spatial coordinates $(x, y)$ of receiver points as input and generates a 16D environmental feature vector $\mathbf{f}_\text{env} \in \mathbb{R}^{16}$. 
	This vector captures key physical characteristics of the environment, such as building layouts and reflective surfaces, so as to guide the learning in subsequent modules.

	\textit{\textbf{Main Network}}: The main network adopts a deep fully connected architecture, including an input layer, three hidden layers, and an output layer. 
	It maps 2D coordinates to a 256D feature space, and then combines these with the 16D scene features from the environment encoder. The output is an $L \times 4$D vector representing the power, delay, elevation, and azimuth of $L$ paths.
	This network models the complex mapping from spatial coordinates to multipath parameters through deep nonlinear transformations and residual learning.	

	\textit{\textbf{Path Classification Layer}}: This fully connected layer takes the combined features from the environment encoder and main network as input and outputs an $L \times 4$ probability matrix, representing the probability distribution over four types of propagation paths: LoS, reflection, scattering, and diffraction. 

	PINN employs a composite loss function comprising supervised, physical, and path-type classification losses, with dynamic weights to balance their contributions, effectively integrating data-driven learning with physical constraints.

	\textbf{Supervised Loss}: The supervised loss ensures accurate fitting to the measurement data, defined as
	\vspace{-0.1cm}
	\begin{equation}
		\mbox{\small$\displaystyle
		\mathcal{L}_\text{sup} = \frac{1}{N}\sum_{i=1}^{N}\left\|\textbf{y}_\text{pred}^i - \textbf{y}_\text{true}^i\right\|_2^2,
		$}
		\vspace{-0.15cm}
	\end{equation}
	where $N$ is the number of training samples, 
	$\textbf{y}_\text{pred}^i$ and $\textbf{y}_\text{true}^i$ denote the predicted and true values for the $i$-th sampling, respectively.

	\textbf{Physical Loss}: 
	The complete physics loss includes four parts: power constraint, delay constraint, angular constraint, and consistency constraint.

	\textit{Power constraint}: The theoretical LoS path is modeled using free-space propagation, where the path loss is given by
    \begin{equation}
		\mbox{\small$\displaystyle
        PL(\text{dB}) = 20\log_{10}(d)+20\log_{10}(f_c)+20\log_{10}\left(\frac{4\pi}{c}\right),
		$}
		\vspace{-0.15cm}
    \end{equation}
	where $d$ is the distance between the transmitter and receiver.

	The predicted power of the LoS path is set to exactly match the theoretical power $p_{r,\text{theory}}^\text{LoS}$. The maximum theoretical power values for reflection, scattering, and diffraction paths are set to 0.7, 0.3, and 0.1 times $p_{r,\text{theory}}^\text{LoS}$, respectively. Accordingly, the power attenuation constraint is formulated as
	\vspace{-0.1cm}
	\begin{equation}
		\mbox{\small$\displaystyle
        \mathcal{L}_{\text{power}} = \sum_{l=1}^L w_{l,\text{type}}^\text{power} \cdot \left(p_\text{pred}^l - {p}_\text{theory}^l\right)^2,
		$}
		\vspace{-0.1cm}
    \end{equation}
	where $w_{l,\text{type}}^\text{power}$ denotes the weight of power associated with the $l$-th path's type, $p_{\text{pred}}^l$ and $p_{\text{theory}}^l$ represent the predicted and theoretical power of the $l$-th path.

	% \addtolength{\topmargin}{0.017in}

	\textit{Delay constraint}: The delay constraint is based on the consideration that the RF signal delay is proportional to the distance. The theoretical LoS delay is given by $\tau_\text{theory}^\text{LoS} = {d}/{c}$.
    The delay of the LoS path is set to be equal to that, and then the minimum delays for reflected, scattered, and diffracted paths are set to 1.2, 1.4, and 1.6 times $\tau_{\text{theory}}^\text{LoS}$, respectively. The delay constraint is expressed as
	\vspace{-0.12cm}
	\begin{equation}
		\mbox{\small$\displaystyle
        \mathcal{L}_{\text{delay}} = \sum_{l=1}^L w_{l,\text{type}}^\text{delay} \cdot \left[\text{ReLU}\left(\tau_\text{theory}^l - {\tau}_\text{pred}^l\right)\right]^2,
		$}
		\vspace{-0.12cm}
    \end{equation}
	where $w_{l,\text{type}}^\text{delay}$ denotes the weight of delay, the $\text{ReLU}(\cdot)$ function ensures that only delays shorter than the theoretical minimum are penalized.

	\textit{Angular constraint}: 
	Based on the geometric relationships of RF signal arrival angles, the theoretical LoS angles are calculated and denoted as $\theta_\text{theory}^\text{LoS}$ and $\varphi_\text{theory}^\text{LoS}$.
	The angular differences are computed as $\Delta \theta = \left|\theta_\text{pred}^l - \theta_\text{theory}^\text{LoS}\right|$ and 
	\vspace{-0.1cm}
    \begin{equation}
		\mbox{\small$\displaystyle
        \Delta \varphi = \min \left(\left|\varphi_\text{pred}^l - \varphi_\text{theory}^\text{LoS}\right|, 360^{\circ} - \left|\varphi_\text{pred}^l - \varphi_\text{theory}^\text{LoS}\right| \right).
		$}
    \end{equation}
	The LoS angular error is defined as $E_l^{\text{LoS}} = \Delta \theta + \Delta \varphi$. For NLoS paths such as reflections, deviations from theoretical angles are allowed due to environmental complexity. For reflected paths, the acceptable deviations in elevation and azimuth angles are $45^\circ$ and $90^\circ$, respectively, and are modulated by the reflection characteristic factor $r_l$ output from the environment encoder. The angular error for a reflected path is computed as
	\vspace{-0.15cm}
	\begin{equation}
		\mbox{\small$\displaystyle
        E_l^\text{refl} = \max\left(0, \Delta \theta-45^\circ \cdot r_l\right)^2 + \max\left(0, \Delta \varphi-90^\circ \cdot r_l\right)^2.
		$}
		\vspace{-0.08cm}
    \end{equation}
	The overall angular constraint is given by
	\vspace{-0.15cm}
	\begin{equation}
		\mbox{\small$\displaystyle
        \mathcal{L}_{\text{angle}} = \sum_{l=1}^L w_{l,\text{type}}^\text{angle} \cdot E_l^{type},
		$}
		\vspace{-0.12cm}
    \end{equation}
	where $w_{l,\text{type}}^\text{angle}$ denotes the weight of angle, and $type$ denotes the path type, including LoS, reflection, scattering, and diffraction.

	\textit{Consistency constraint}: The consistency constraint $\mathcal{L}_{\text{consist}}$ is designed to ensure the physical coherence among multipath parameters. Specifically, it imposes a logical relationship postulating that the received power at each point is inversely related to the delay. Meanwhile, based on the predicted power and delay, the probability of the existence of a LoS path, denoted as $p_\text{LoS}$, is inferred.

	In summary, the physical loss is expressed as
	\vspace{-0.15cm}
		\begin{equation}
		\vspace{-0.1cm}
		\mbox{\footnotesize$\displaystyle
        \mathcal{L}_\text{phy} =\frac{1}{N}\sum_{i=1}^{N} \left(\lambda_1 \mathcal{L}_{\text{power}}^i + \lambda_2\mathcal{L}_{\text{delay}}^i + \lambda_3\mathcal{L}_{\text{angle}}^i + \lambda_4\mathcal{L}_{\text{consist}}^i\right),
		$}
    \end{equation}
	where $\lambda_j,j \in \left\{1,2,3,4\right\}$ are the weights assigned to the four components, respectively, and $\mathcal{L}^i$ is the loss for the $i$-th sample.

	\textbf{Path-type Classification Loss}: The path-type classification loss $\mathcal{L}_{\text{type}}$ is employed to regulate the output of the PINN's path classifier, ensuring consistency in the model's predictions across different path types.  This loss is formulated as a cross-entropy function
	\vspace{-0.15cm}
	\begin{equation}
		\mbox{\small$\displaystyle
        \mathcal{L}_{\text{type}} = -\frac{1}{N}\sum_{i=1}^{N} \sum_{l=1}^{L} \sum_{k=1}^{4} y_{i,l,k} \log(p_{i,l,k}),
		$}
		\vspace{-0.1cm}
    \end{equation}
	where $k$ denotes the path type index, $y_{i,l,k}$ is the true label of the $l$-th path for the $i$-th sample which is constructed based on physical rules. 

	Accordingly, the composite loss function of the PINN is expressed as
	\vspace{-0.15cm}
	\begin{equation}
		\mbox{\small$\displaystyle
        \mathcal{L}_\text{PINN} = \mathcal{L}_\text{sup} + \lambda_\text{phy} \mathcal{L}_\text{phy} + \lambda_\text{type} \mathcal{L}_\text{type}, 
		$}
		\vspace{-0.1cm}
    \end{equation}
	where $\lambda_\text{phy}$ and $\lambda_\text{type}$ are dynamic weighting coefficients, gradually increased from 0.05 to 0.5 and 0.3 during training.
	\vspace{-0.1cm}

    \subsection{GNN Architecture}

	In the PINN stage, multipath parameters at each receiver are predicted independently, without enforcing spatial consistency. To overcome this limitation, a GNN is introduced to model spatial correlations among receivers, refine PINN outputs, and enhance prediction accuracy.

	The GNN takes as input a coordinate matrix $\mathbf{X} \in \mathbb{R}^{N \times 2}$ and the initial multipath feature matrix $\mathbf{Y}^{(0)} \in \mathbb{R}^{N \times (4*L)}$ predicted by the PINN, where each node encodes four parameters across $L$ paths. The network outputs an optimized feature matrix $\mathbf{Y}^{(\text{out})} \in \mathbb{R}^{N \times (4*L)}$. A graph $\mathcal{G}=(\mathcal{V},\mathcal{E})$ is built using the k-nearest neighbor algorithm, where $\mathcal{V}$ denotes the set of nodes and $\mathcal{E}$ the set of edges.

	The network architecture consists of four GraphSAGE convolutional layers: one input layer, two 128D hidden layers, and one output layer. Each layer employs mean aggregation and skip connections. During training, the MSE loss function is used to minimize the difference between the predicted and true multipath parameters, which is defined as
	\vspace{-0.13cm}
	\begin{equation}
		\mbox{\small$\displaystyle
        \mathcal{L}_\text{GNN}=\frac{1}{M} \sum_{j=1}^M \left\| \mathbf{y}^j_{\text{out}}-\mathbf{y}^j_\text{true}\right\| _2^2,
		$}
		\vspace{-0.15cm}
    \end{equation}
	where $M$ is the number of training samples, $\mathbf{y}^j_{\text{out}}$ and $\mathbf{y}^j_\text{true}$ denote the predicted and true multipath parameters for the $j$-th sample, respectively.

	\subsection{Joint Construction of RF Map via PINN and GNN}

	As illustrated in Fig. \ref{fig:PINN_GNN}, the proposed framework first employs a PINN to coarsely estimate multipath parameters from sparse samples while enforcing physical consistency via embedded constraints. A GNN then refines these predictions by modeling spatial correlations across receiver locations, enhancing generalization. The full construction process is summarized in Algorithm \ref{alg:PINN_GNN}. By integrating physical priors and spatial dependencies, the framework achieves accurate multipath parameter estimation.

	\vspace{-0.31cm}
	\begin{algorithm}
		\caption{RF Map construction Algorithm with Joint PINN and GNN}
		\label{alg:PINN_GNN}
		\KwIn{Coordinates of points $\mathbf{X} \in \mathbb{R}^{N \times 2}$, multipath data of known points $\mathbf{Y}_\text{known} \in \mathbb{R}^{M \times (4*L)}$}
		\KwOut{Complete RF map $\mathbf{\hat{Y}} \in \mathbb{R}^{N \times (4*L)}$}
	
		Data preprocessing and feature-level normalization\;

		\tcp{\textbf{PINN Training}}
		\For{each epoch}{
		Forward process: $\mathbf{y}_\text{PINN} = \text{PINN}(\mathbf{x}_{\text{known}})$\;
		Calculate $\mathcal{L}_\text{PINN} = \mathcal{L}_{\text{MSE}} + \lambda_{\text{phy}}\mathcal{L}_{\text{phy}} + \lambda_{\text{type}}\mathcal{L}_{\text{type}}$\;
		Backward process to update PINN parameters\;
		}

		\tcp{\textbf{GNN Training}}
		Construct k-nearest neighbor graph $\mathcal{G}=(\mathcal{V},\mathcal{E})$\;
		\For{each epoch}{
		Forward propagation: $\mathbf{y}_\text{GNN} = \text{GNN}(\mathbf{y}_\text{PINN}, \mathcal{G})$\;
		Calculate $\mathcal{L}_\text{GNN}$\;
		Update GNN parameters\;
		}

		\tcp{\textbf{RF Map Construction}}
		Initial prediction by PINN: $\hat{\mathbf{y}}_\text{PINN} = \text{PINN}(\mathbf{x}_\text{all})$\;
		Construct complete graph structure $\mathcal{G}_\text{full}$\;
		Optimize results by GNN: $\hat{\mathbf{y}}_\text{GNN} = \text{GNN}(\hat{\mathbf{y}}_\text{PINN}, \mathcal{G}_\text{full})$\;
		Post-processing and inverse normalization\;
		\Return Complete RF map $\mathbf{\hat{Y}} \in \mathbb{R}^{N \times (4*L)}$.\\
	\end{algorithm}
	\vspace{-0.4cm}

	\section{Simulation and Performance Evaluation}
	
	\subsection{Simulation Scenario and Setup}

\begin{figure}[htbp]
		\centering
		\subfigure[] {\label{DeepMIMO}\centering\includegraphics[width=0.32\columnwidth]{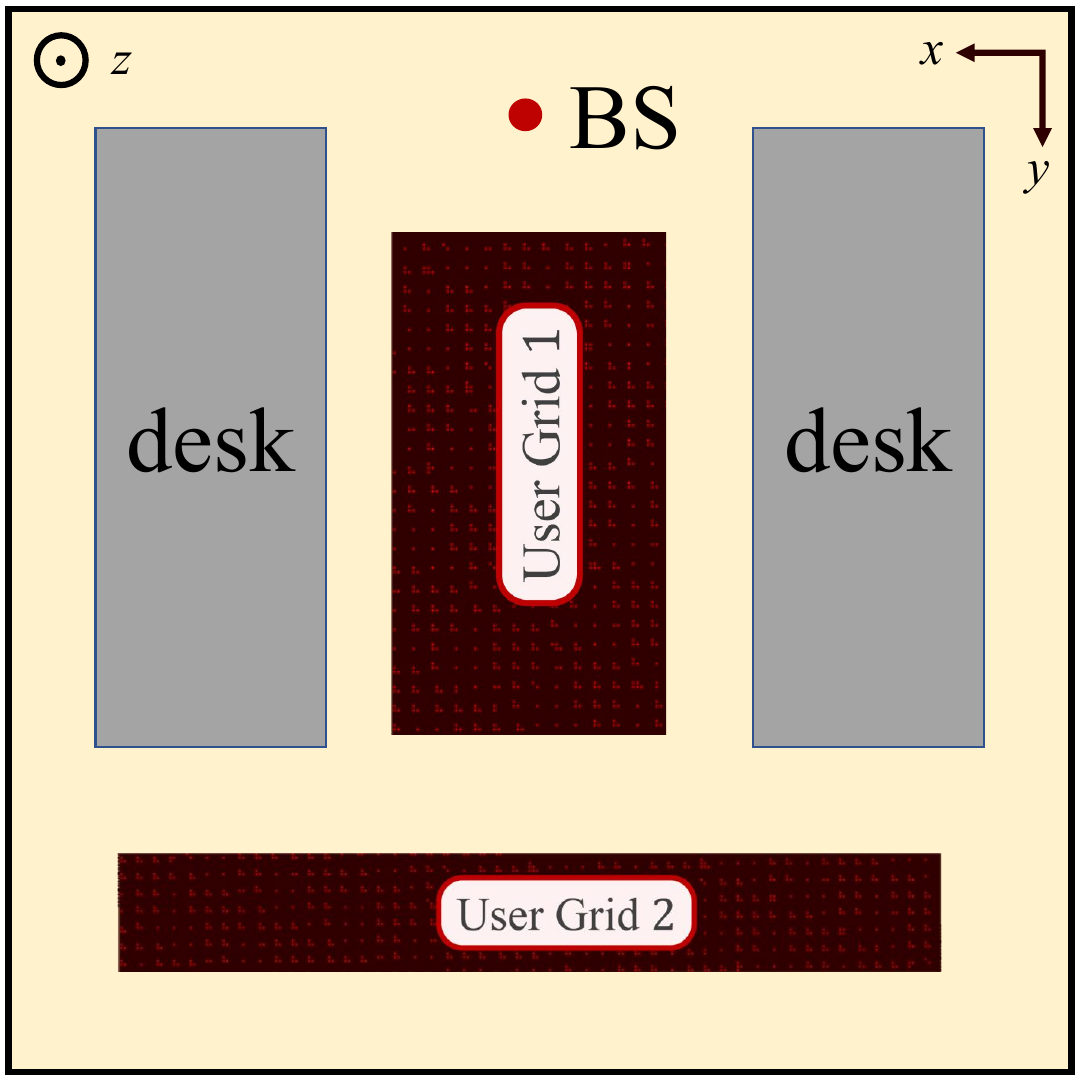}%
		}
		% \vspace{-0.25cm}
		\subfigure[] {\label{USTC1}\centering\includegraphics[width=0.32\columnwidth]{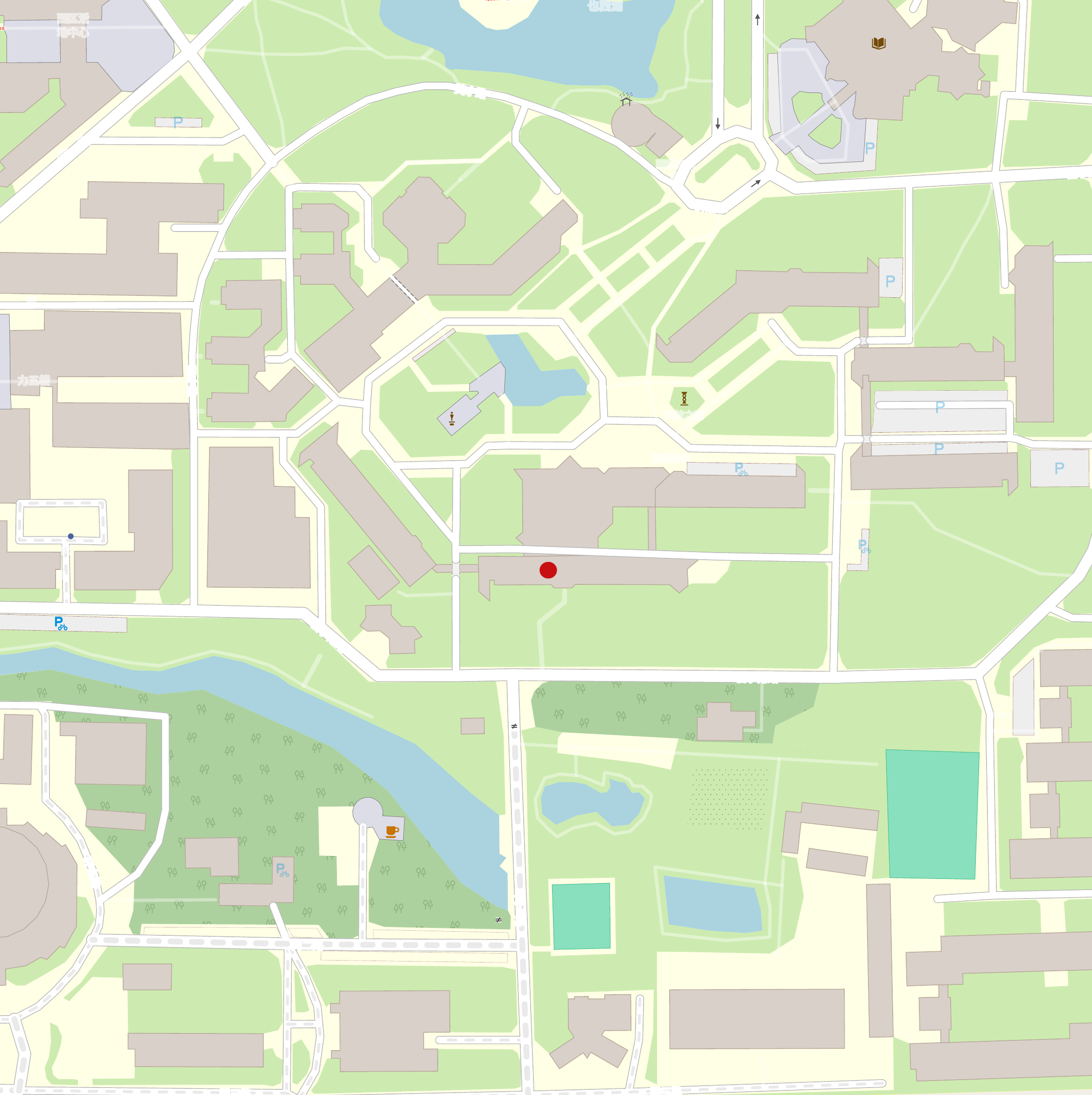}%
		}
		\subfigure[] {\label{USTC2}\centering\includegraphics[width=0.32\columnwidth]{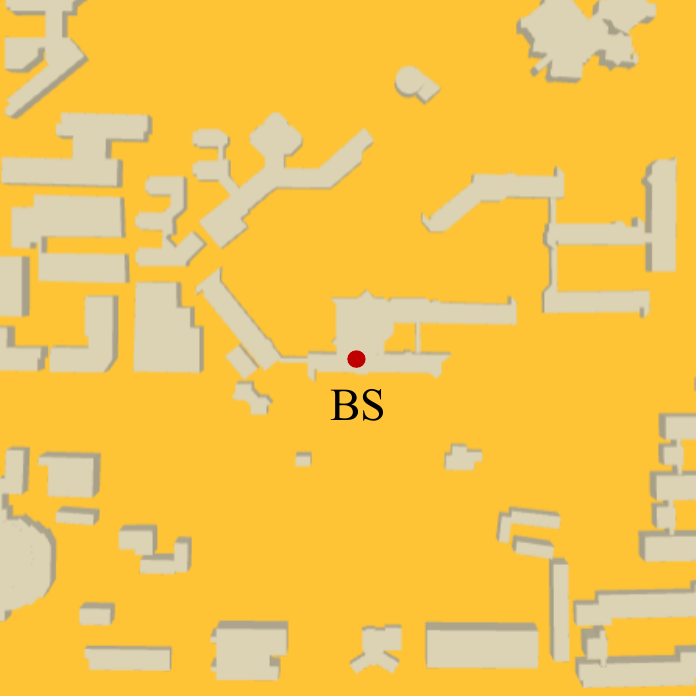}%
		}
		\vspace{-0.25cm}
		\caption{Simulation scenarios: (a) DeepMIMO dataset, (b) USTC campus environment, (c) 3D model of USTC campus buildings.}
		\vspace{-0.5cm}
		\label{fig:scenarios}
	\end{figure}

	Experiments were conducted in two representative scenarios. The indoor scenario (S1), based on the DeepMIMO dataset \cite{DeepMIMO}, as shown in Fig. \ref{DeepMIMO}, features a $10 \times 10 \times 5~\text{m}^3$ space with the access point (AP) at 2.5 m and 30,280 receivers placed every 50 cm across two grids, operating at 2.4 GHz. The outdoor scenario (S2) covers a $512 \times 512~\text{m}^2$ area on the West Campus of USTC, as illustrated in Fig. \ref{USTC1}, with 3D-modeled buildings (Fig. \ref{USTC2}) \cite{WNet} and multipath data generated via Sionna ray tracing \cite{Sionna}. The ray-traced multipath data serve as the ground truth for subsequent model training and performance evaluation. Receivers are spaced 1 m apart and elevated 1 m above ground. The base station (BS) is placed 30 m high above the Third Electronics Building, operating at 3.5 GHz. Since the fifth path's power drops to 2\%-5\% of the first, the number of multipath components is set to $L=5$.

	Four sampling rates (5\%, 10\%, 15\%, 20\%) are evaluated, where multipath data at the aforementioned randomly selected sparse points are known. Known data is split into training, validation, and testing sets in a 7:1.5:1.5 ratio. The model is trained with five rounds of data augmentation. The validation set is used for tuning, and the test set evaluates performance before inference. All performance metrics reported in the results are computed based on the prediction errors at unknown points concerning the true values from ray tracing. PINN and GNN are trained for up to 300 and 800 iterations using Adam (initial learning rate 0.001) with cosine annealing, weight decay of $1\text{e}{-5}$, and dropout rate of 0.1. Training and inference are conducted on two NVIDIA V100 GPUs.

	\vspace{-0.25cm}

	\subsection{Simulation Results}

	To comprehensively validate the effectiveness of the proposed PINN-GNN framework, we compare it with four baseline methods: Kriging, Separate, GNN\_only, and PINN\_only.
	Kriging is a spatial interpolation technique based on Gaussian process regression \cite{Kriging}. The Separate method also adopts our proposed framework but treats each multipath parameter separately, ignoring their correlations. GNN\_only and PINN\_only use GNN or PINN alone for prediction.
	To evaluate performance, we adopt MSE, root MSE (RMSE), and normalized MSE (NMSE) as metrics.

	\begin{table*}[htbp]
		\centering
		\setlength{\tabcolsep}{4pt} % Reduced column spacing from 5pt to 4pt
		\renewcommand{\arraystretch}{0.8} % Reducing row height by setting to less than 1
		\caption{Performance Metrics at Different Sparse Sampling Rates}
		\vspace{-0.2cm}
		\small
		\begin{tabular}{l|cc|cc|cc|cc}
		\toprule
		 & \multicolumn{2}{c|}{5\%} & \multicolumn{2}{c|}{10\%} & \multicolumn{2}{c|}{15\%} & \multicolumn{2}{c}{20\%} \\
		\cmidrule(lr){2-3} \cmidrule(lr){4-5} \cmidrule(lr){6-7} \cmidrule(lr){8-9}
		 & S1 & S2 & S1 & S2 & S1 & S2 & S1 & S2 \\
		\midrule
		MSE & 121.76419 & 587.27797 & 82.72304 & 330.99339 & 73.86910 & 256.36348 & 66.09599 & 203.52467 \\
		RMSE & 11.03468 & 24.23382 & 9.09522 & 18.19322 & 8.59471 & 16.01135 & 8.12994 & 14.26621 \\
		NMSE & 0.02663 & 0.14302 & 0.01808 & 0.07242 & 0.01608 & 0.06297 & 0.01446 & 0.05025 \\
		\bottomrule
		\end{tabular}
		\label{table1}
		\vspace{-0.45cm}
	\end{table*}

	Table \ref{table1} presents the performance of the proposed PINN-GNN framework under different sampling rates in S1 and S2. Increasing the rate from 5\% to 10\% improves accuracy, reducing MSE by 32\% in S1 and 44\% in S2. However, gains diminish beyond 10\% sampling rate, suggesting a limited benefit from denser sampling. Balancing accuracy and cost, 10\% is chosen for subsequent experiments. Prediction errors in S2 are generally higher due to its more complex outdoor environment, including frequent NLoS conditions, larger receiver spacing, and more intricate building layouts.
	\begin{table*}[htbp]
		\centering
		\setlength{\tabcolsep}{4pt} % Reduced column spacing from 5pt to 4pt
		\renewcommand{\arraystretch}{0.8} % Reducing row height by setting to less than 1
		\caption{Comparison of Different Methods Across S1 and S2 Scenarios}
		\vspace{-0.2cm}
		\small
		\begin{tabular}{l|cc|cc|cc|cc|cc}
		\toprule
		 & \multicolumn{2}{c|}{Kriging} & \multicolumn{2}{c|}{Separate} & \multicolumn{2}{c|}{GNN\_only} & \multicolumn{2}{c|}{PINN\_only} & \multicolumn{2}{c}{Proposed} \\
		\cmidrule(lr){2-3} \cmidrule(lr){4-5} \cmidrule(lr){6-7} \cmidrule(lr){8-9} \cmidrule(lr){10-11}
		 & S1 & S2 & S1 & S2 & S1 & S2 & S1 & S2 & S1 & S2 \\
		\midrule
		MSE & 159.0799 & 887.1504 & 134.4780 & 682.7204 & 113.5289 & 573.1187 & 101.5289 & 512.6815 & \textbf{82.7230} & \textbf{330.9934} \\
		RMSE & 12.6127 & 29.7851 & 11.5965 & 26.1289 & 10.6550 & 23.9399 & 10.0762 & 22.6425 & \textbf{9.0952} & \textbf{18.1932} \\
		NMSE & 0.0347 & 0.2172 & 0.0295 & 0.1681 & 0.0248 & 0.1397 & 0.0223 & 0.1281 & \textbf{0.0181} & \textbf{0.0724} \\
		\bottomrule
		\end{tabular}
		\label{table2}
		\vspace{-0.45cm}
	\end{table*}

	\begin{figure*}[htbp]
		\centering
		\subfigure{\label{fig:heatmap1}\centering\includegraphics[width=0.21\textwidth]{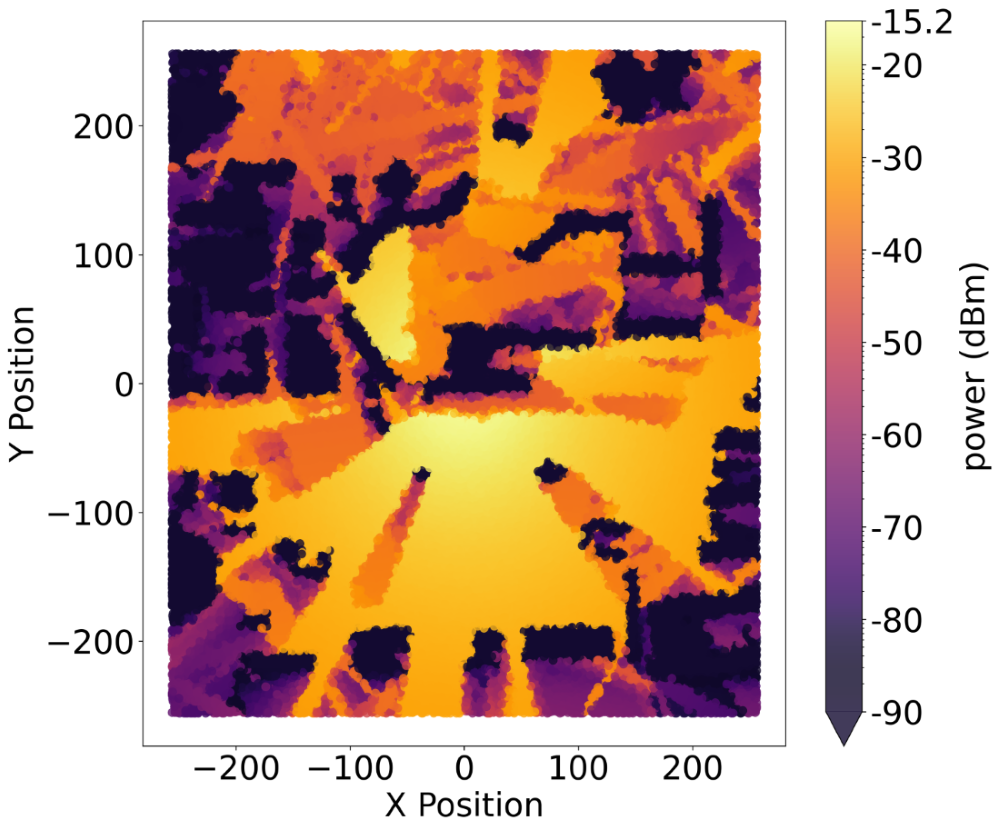}%
		}
		\subfigure{\label{fig:heatmap2}\centering\includegraphics[width=0.21\textwidth]{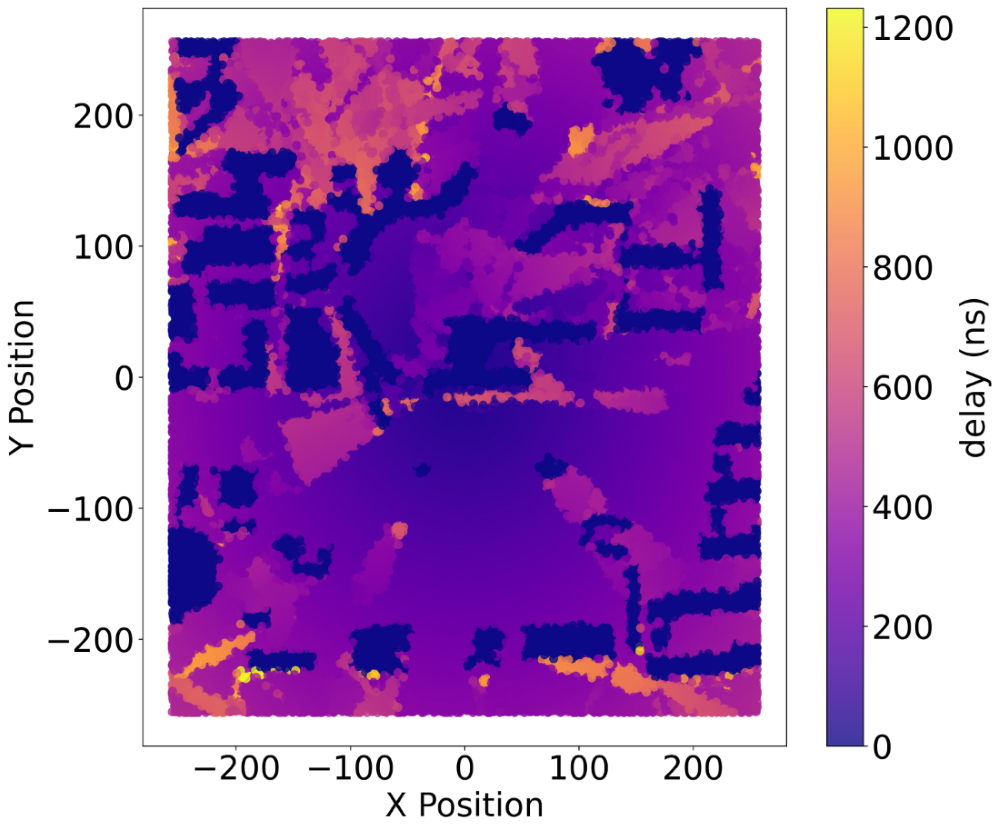}%
		}
		\subfigure{\label{fig:heatmap3}\centering\includegraphics[width=0.21\textwidth]{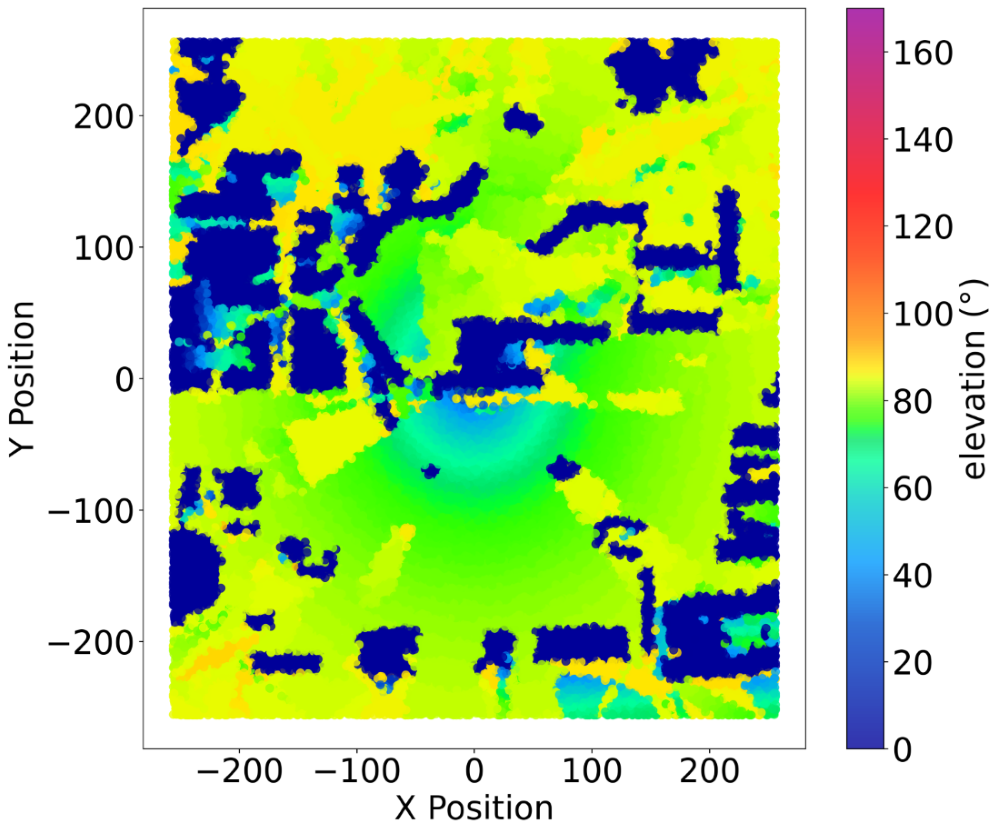}%
		}
		\subfigure{\label{fig:heatmap4}\centering\includegraphics[width=0.21\textwidth]{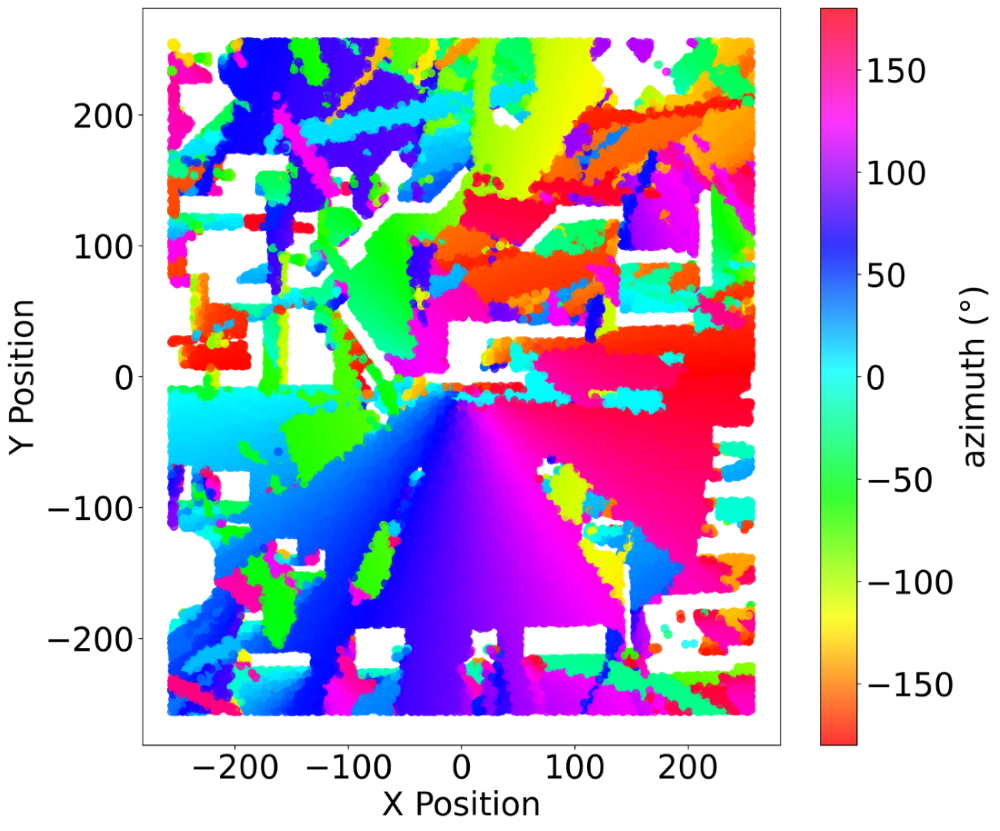}%
		}\\
		\vspace{-0.25cm}
		\subfigure{\label{fig:heatmap5}\centering\includegraphics[width=0.21\textwidth]{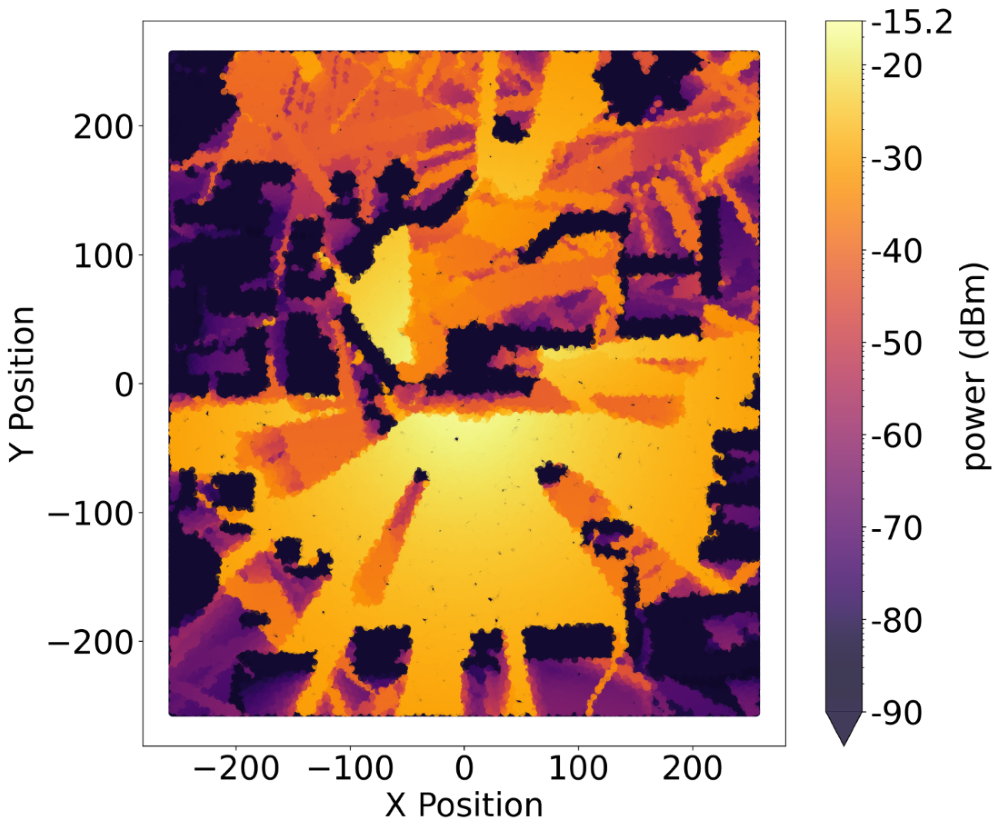}%
		}
		\subfigure{\label{fig:heatmap6}\centering\includegraphics[width=0.21\textwidth]{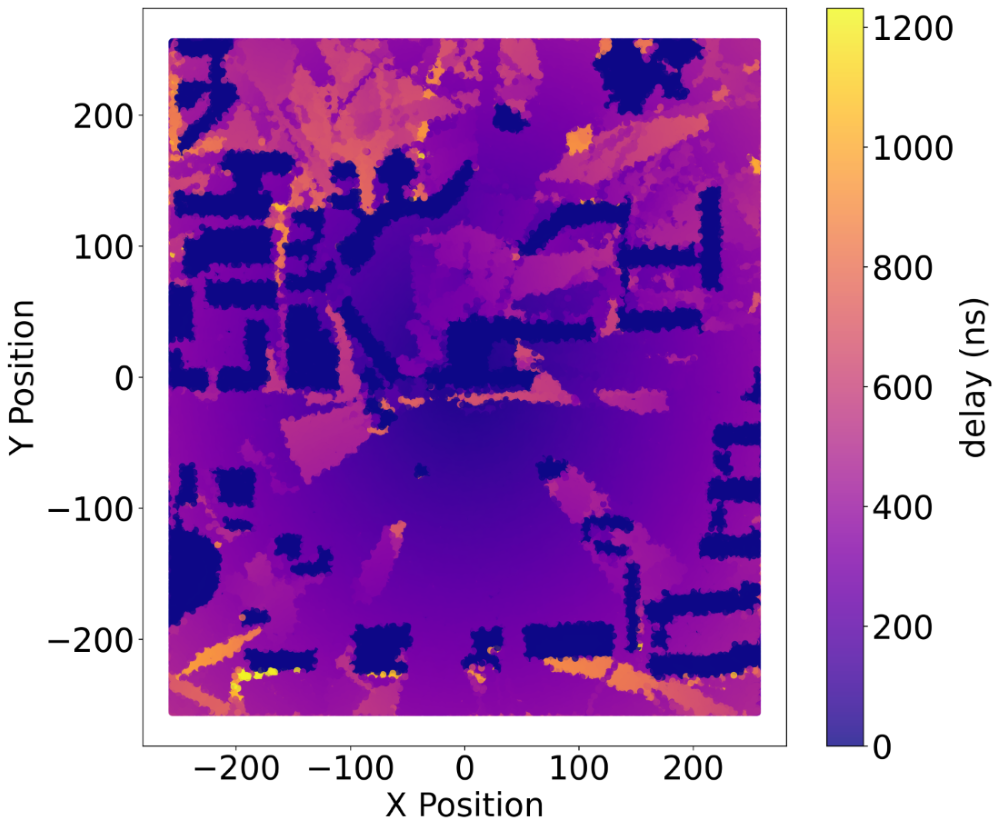}%
		}
		\subfigure{\label{fig:heatmap7}\centering\includegraphics[width=0.21\textwidth]{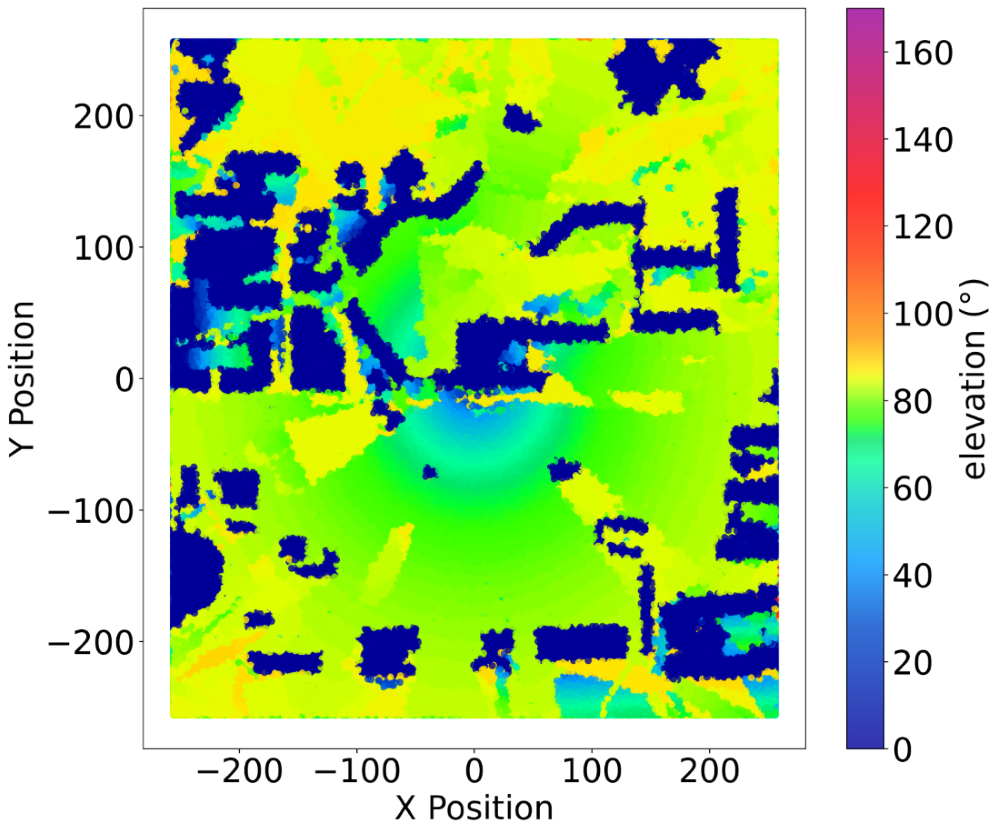}%
		}
		\subfigure{\label{fig:heatmap8}\centering\includegraphics[width=0.21\textwidth]{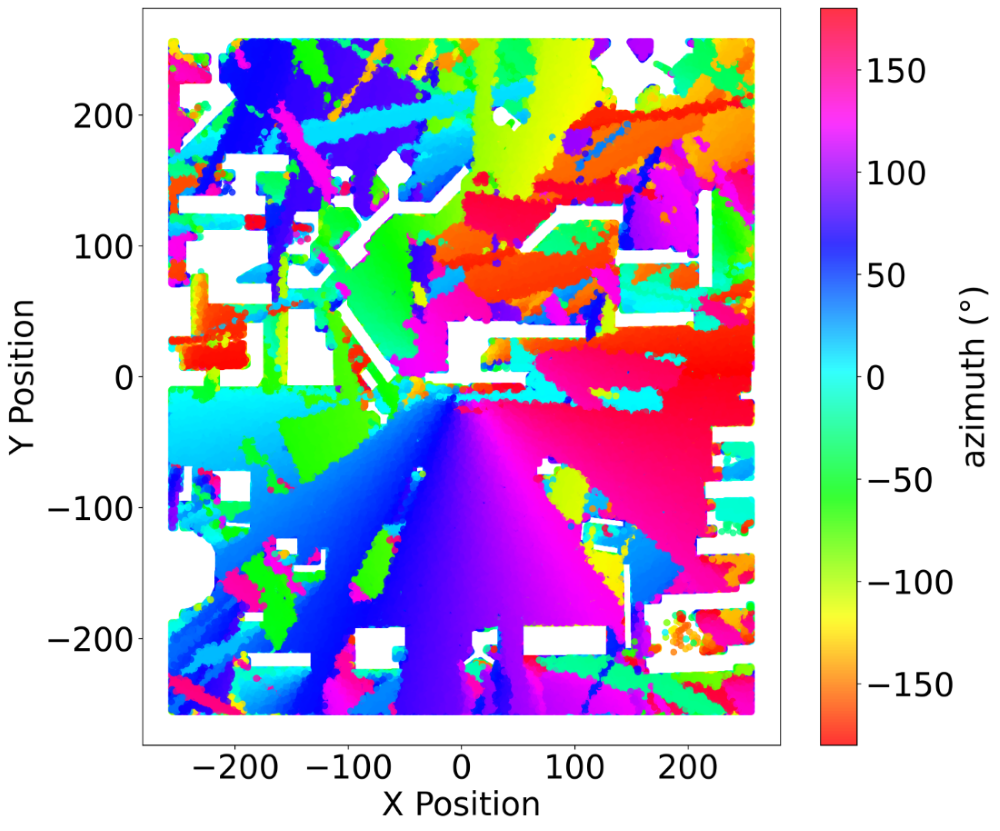}%
		}
		\vspace{-0.25cm}
		\caption{Comparison of true and predicted distributions of the first-path parameters in S2: The first row illustrates the true power, delay, elevation angle, and azimuth angle, respectively, while the second row shows the corresponding predicted results.}
		\vspace{-0.5cm}
		\label{fig:heatmaps}
	\end{figure*}

	Table \ref{table2} compares the proposed PINN-GNN framework with several baselines under 10\% sampling rate in both S1 and S2. The results show clear advantages: MSE is reduced by 48.0\% in S1 and 62.7\% in S2 compared to Kriging. Against PINN\_only, GNN\_only, and Separate methods, the proposed approach achieves 18.5\%-38.5\% MSE reduction in S1 and 35.4\%-51.5\% in S2. These improvements highlight the benefits of joint prediction, as well as the complementary strengths of integrating physical constraints via PINN and spatial dependencies via GNN.

	\begin{figure*}[htbp]
		\centering
		\subfigure[] {\label{fig:CDF1}\centering\includegraphics[width=0.403\textwidth]{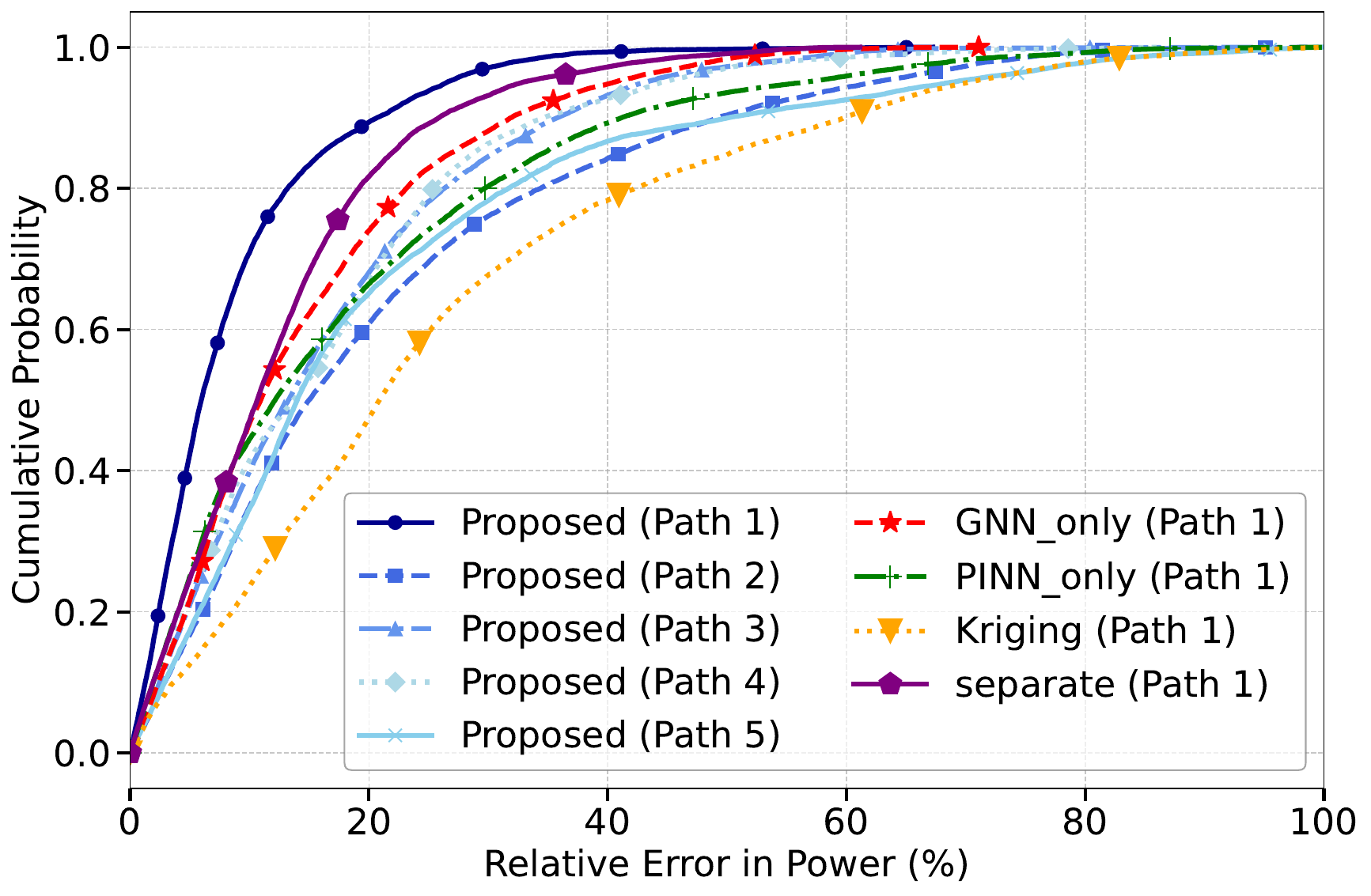}%
		}
		\subfigure[] {\label{fig:CDF2}\centering\includegraphics[width=0.403\textwidth]{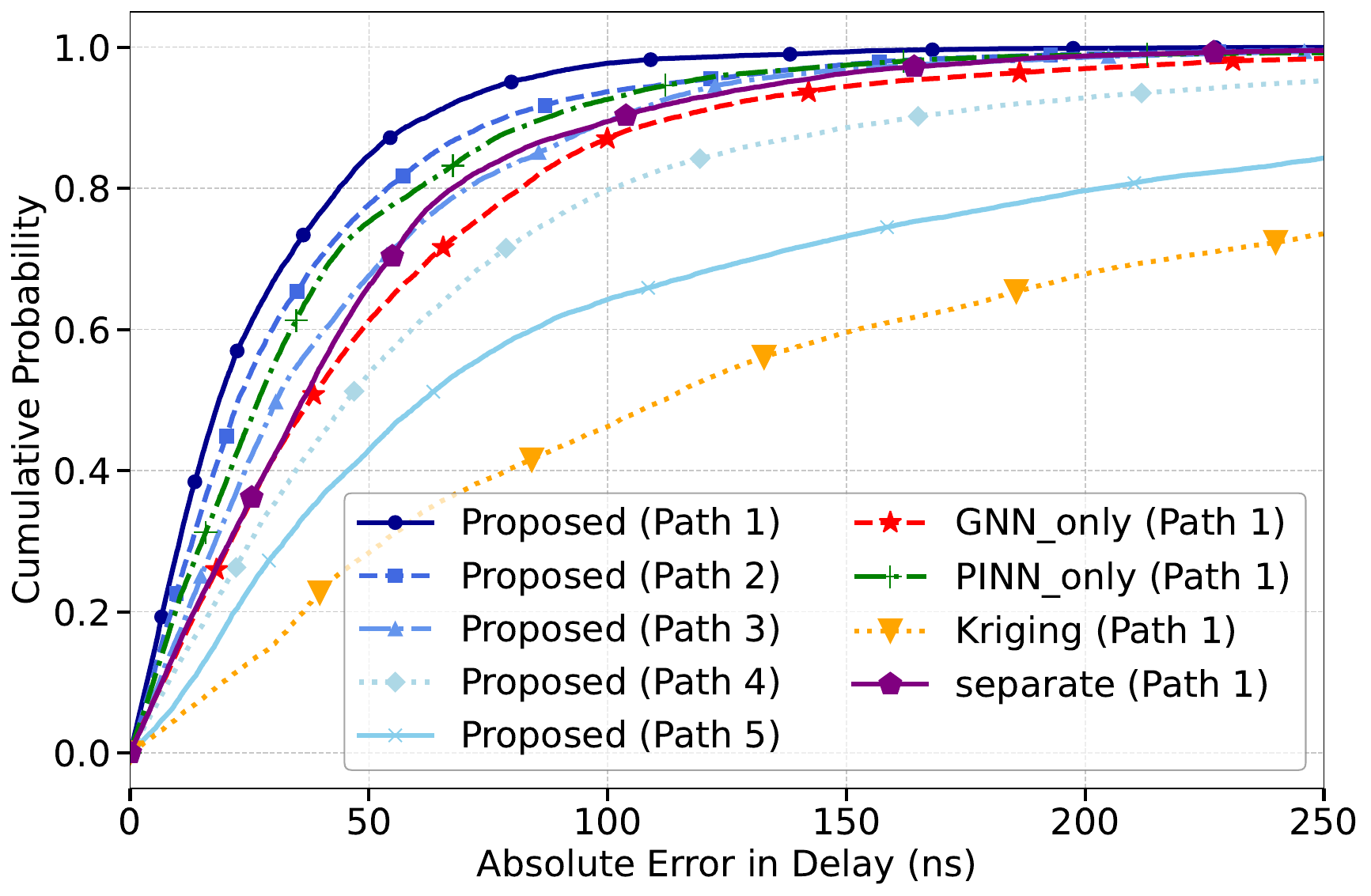}%
		}\\
		\vspace{-0.25cm}
		\subfigure[] {\label{fig:CDF3}\centering\includegraphics[width=0.403\textwidth]{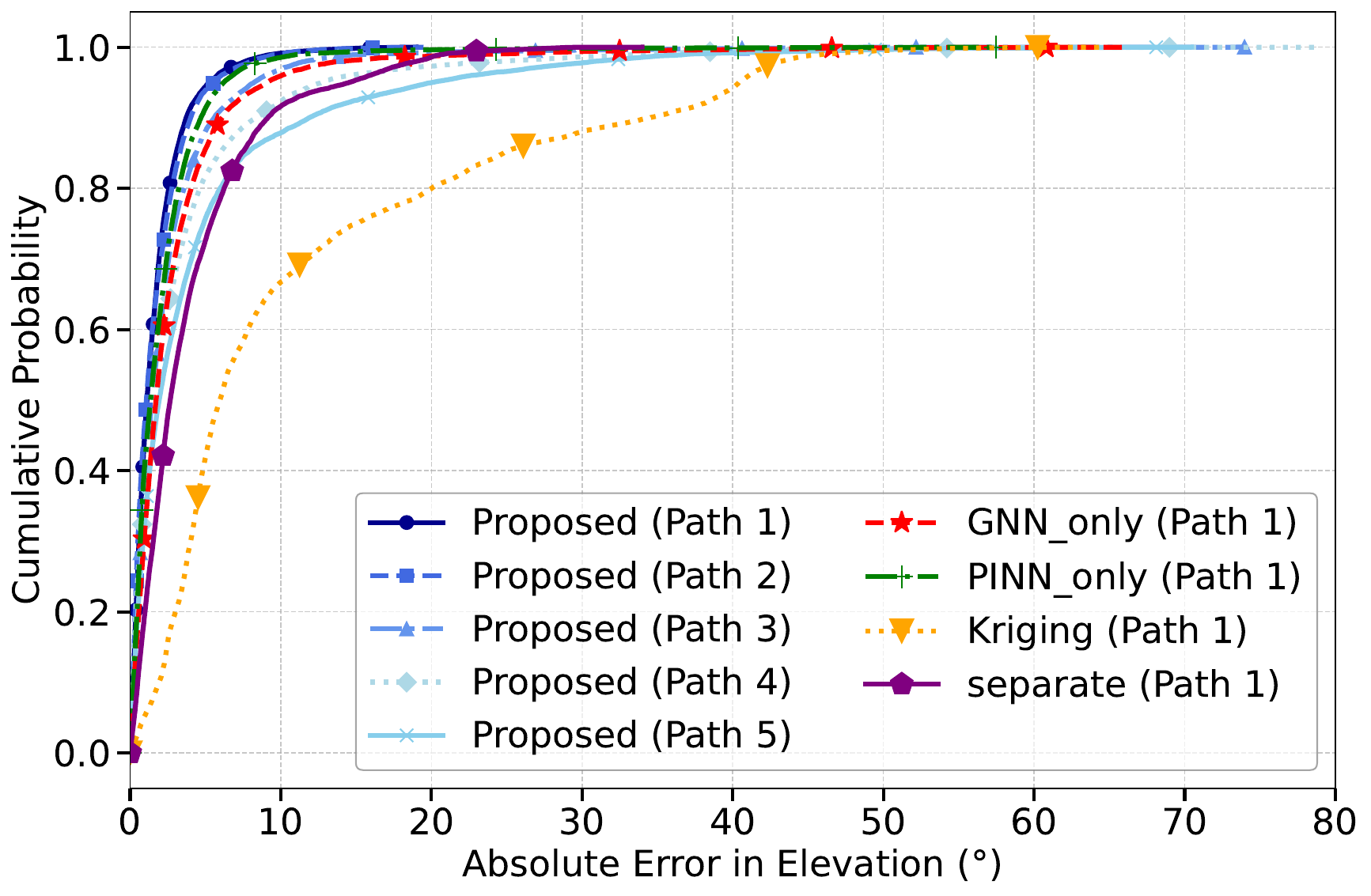}%
		}
		\subfigure[] {\label{fig:CDF4}\centering\includegraphics[width=0.403\textwidth]{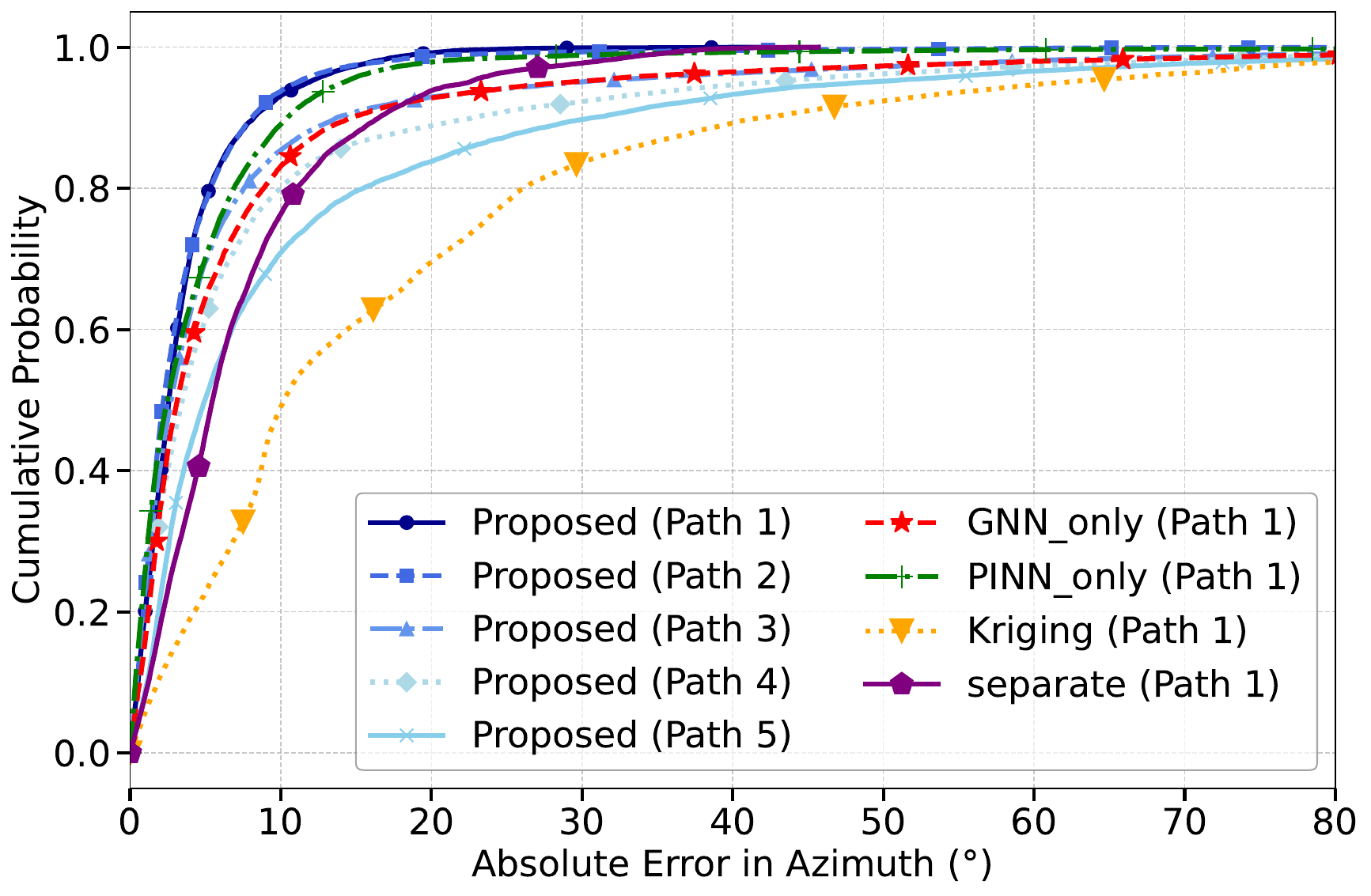}%
		}
		\vspace{-0.25cm}
		\caption{CDF of the errors for the four parameters under different methods in S2: (a) power, (b) delay, (c) elevation angle, and (d) azimuth angle.}
		\vspace{-0.5cm}
		\label{fig:CDF}
	\end{figure*}

	\begin{figure}[htbp]
		\centering
		\subfigure[] {\label{fig:cir1}\centering\includegraphics[width=0.94\columnwidth]{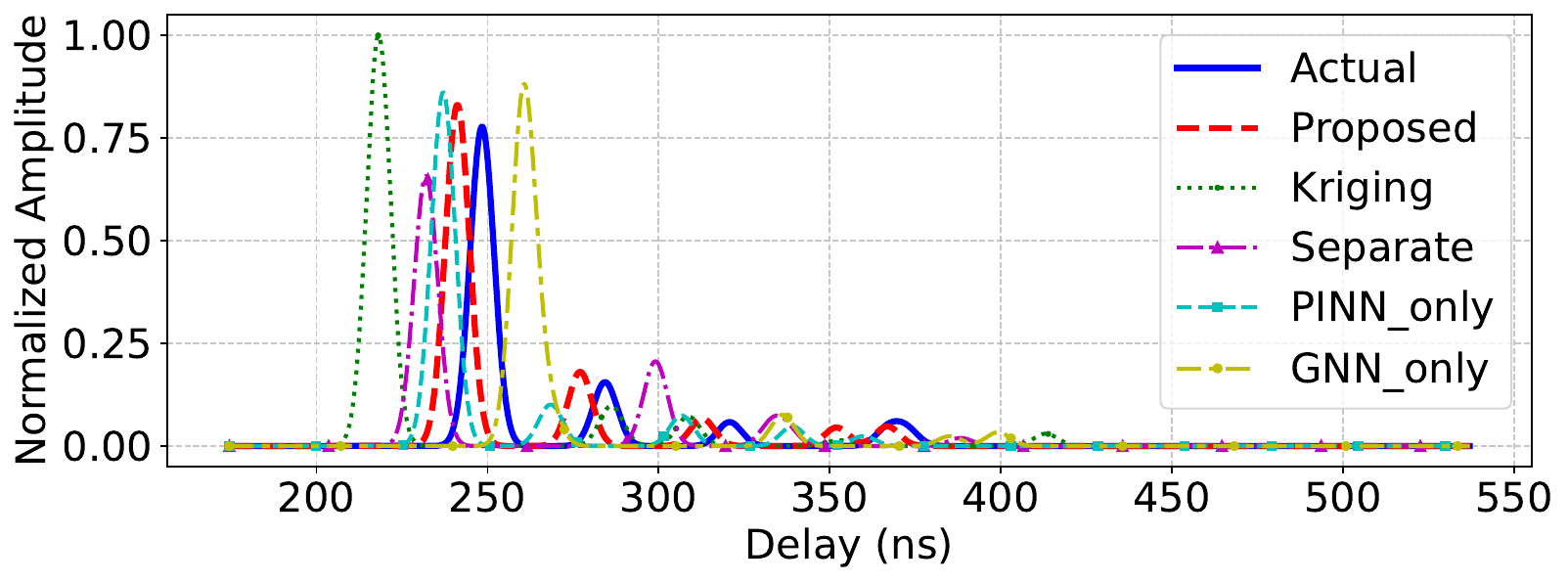}%
		}\\
		\vspace{-0.25cm}
		\subfigure[] {\label{fig:cir2}\centering\includegraphics[width=0.94\columnwidth]{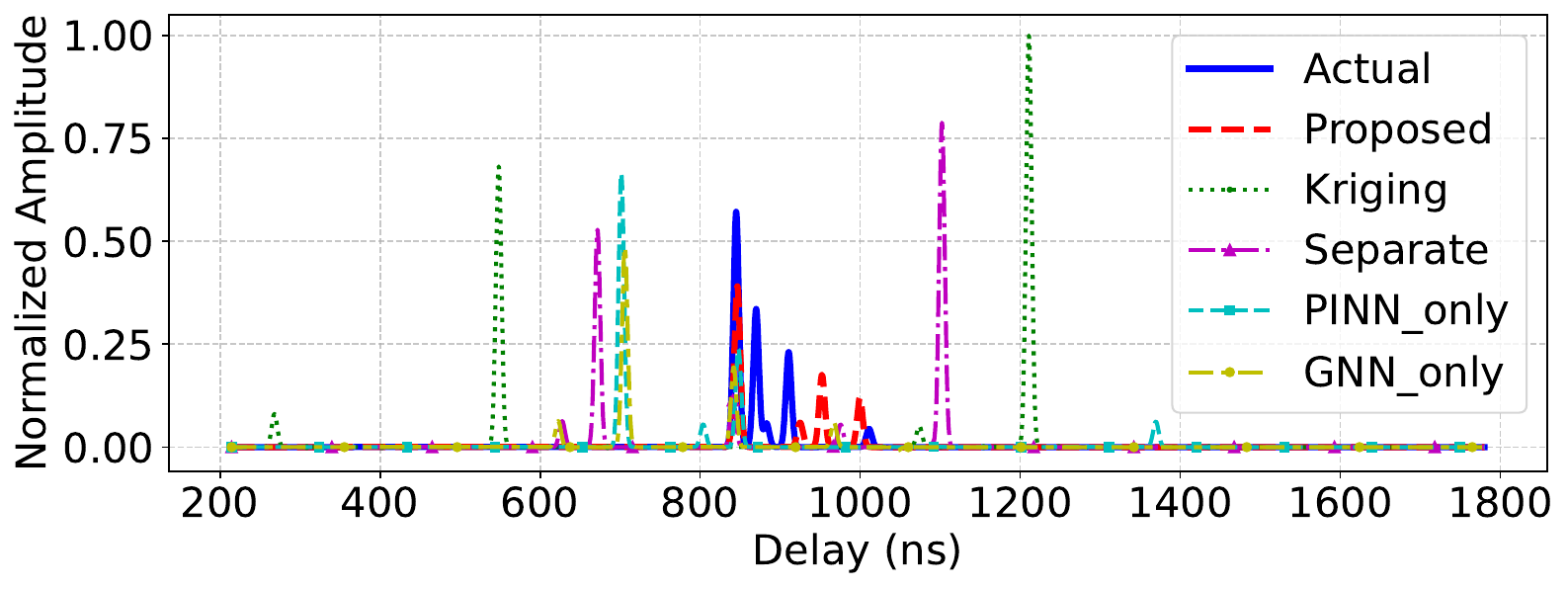}%
		}
		\vspace{-0.25cm}
		\caption{Channel impulse responses under different methods: (a) at location (-18.59, 18.67, 1) in S1, and (b) at location (160.52, 187.27, 1) in S2.}
		\vspace{-0.4cm}
		\label{fig:cir}
		\vspace{-0.45cm}
	\end{figure}

	Fig. \ref{fig:heatmaps} compares the predicted and true values of four key first-path parameters in S2. The predicted power, delay, and elevation maps accurately reflect the radial pattern centered at the BS, while the azimuth prediction captures the expected circular distribution, highlighting the model's ability to learn physical propagation characteristics. The sharp reconstruction of building outlines further confirms its spatial feature modeling capability. The strong alignment between predicted and true values demonstrates the effectiveness of the PINN-GNN framework in combining physical constraints with spatial correlations for accurate multipath modeling.

	Fig. \ref{fig:CDF} shows the cumulative distribution functions (CDFs) curves of the aforementioned four parameters for five predicted paths using the proposed framework and the first path from the four baseline methods. Results show that PINN-GNN consistently outperforms all baselines. For the first path, 90\% of power errors are below 0.2, most of delay errors within 100 ns, elevation errors under $8^{\circ}$, and over 90\% of azimuth errors below $10^{\circ}$. While errors slightly increase for later paths, all predictions remain within reasonable bounds, confirming the robustness of the PINN-GNN framework in complex environments.

	Fig. \ref{fig:cir} shows the predicted time-domain channel impulse responses (CIR) at random receiver locations in S1 and S2. In the indoor scenario S1, the proposed PINN-GNN framework yields a predicted CIR profile closely matching the true one, with a slight time offset, thanks to stable LoS paths and strong modeling of dominant components. In the more complex outdoor S2, despite stronger multipath, the method maintains high prediction accuracy and outperforms all baselines. The Kriging method shows significant deviations in S2, reflecting its limitations in large, sparsely sampled environments. Overall, the results confirm the robustness and accuracy of the proposed framework across varied propagation scenarios.

	\section{Conclusion}
	
	In this paper, we propose a PINN-GNN framework for RF map construction. The PINN encodes physical constraints to map receiver coordinates to multipath parameters, while the GNN extracts spatial correlations via a k-nearest neighbor graph to enhance prediction. Experiments on the DeepMIMO dataset and a USTC campus scenario show that the framework consistently achieves high accuracy across sparsely sampled environments. Notably, it performs well in complex environments by effectively exploiting the multipath modeling knowledge. Future work will explore its scalability to larger, more challenging scenarios, and its utilization in wireless network systems.

\end{document}